\documentclass{elsart}
\usepackage{ifpdf}
\usepackage{amsfonts}
\usepackage{latexsym}
\usepackage{amssymb}
\usepackage{amsmath}
\usepackage{times}
\usepackage{graphicx,natbib,amssymb,lineno}
\ifpdf
\usepackage[
  pdftitle={Instructions for use of the document class
    elsart},
  pdfauthor={Simon Pepping},
  pdfsubject={The preprint document class elsart},
  pdfkeywords={instructions for use, elsart, document class},
  pdfstartview=FitH,
  bookmarks=true,
  bookmarksopen=true,
  breaklinks=true,
  colorlinks=true,
  linkcolor=blue,anchorcolor=blue,
  citecolor=blue,filecolor=blue,
  menucolor=blue,pagecolor=blue,
  urlcolor=blue]{hyperref}
\else
\usepackage[
  breaklinks=true,
  colorlinks=true,
  linkcolor=blue,anchorcolor=blue,
  citecolor=blue,filecolor=blue,
  menucolor=blue,pagecolor=blue,
  urlcolor=blue]{hyperref}
\fi

\makeatletter
\def\elsartstyle{
    \def\normalsize{\@setfontsize\normalsize\@xiipt{14.5}}
    \def\small{\@setfontsize\small\@xipt{13.6}}
    \let\footnotesize=\small
    \def\large{\@setfontsize\large\@xivpt{18}}
    \def\Large{\@setfontsize\Large\@xviipt{22}}
    \skip\@mpfootins = 18\p@ \@plus 2\p@
    \normalsize
} \@ifundefined{square}{}{} \makeatother

\begin{document}

\begin{frontmatter}



\title{Integrability of the Gross-Pitaevskii Equation with Feshbach Resonance management}
\author[a,*]{Dun Zhao},
\ead{zhaod@lzu.edu.cn}
\thanks[*]{Corresponding author.}
\author[a]{Hua-Yue Chai},
\author[b,c]{Hong-Gang Luo}
\address[a]{School of Mathematics and Statistics, Lanzhou University,
Lanzhou 730000, China}
\address[b]{Center of Theoretical Physics and interdisciplinary Studies,
Lanzhou University, Lanzhou 730000, China}
\address[c]{Institute of Theoretical Physics, Academia Sinica, Beijing 100080, China}

\begin{abstract}
In this paper we study the integrability of a class of
Gross-Pitaevskii equations managed by Feshbach resonance in an
expulsive parabolic external potential. By using WTC test, we find
a condition under which the Gross-Pitaevskii equation is
completely integrable. Under the present model, this integrability
condition is completely consistent with that proposed by Serkin,
Hasegawa, and Belyaeva [V. N. Serkin et al., Phys. Rev. Lett. 98,
074102 (2007)]. Furthermore, this integrability can also be
explicitly shown by a transformation, which can convert the
Gross-Pitaevskii equation into the well-known standard nonlinear
Schr\"odinger equation. By this transformation, each exact
solution of the standard nonlinear Schr\"odinger equation can be
converted into that of the Gross-Pitaevskii equation, which builds
a systematical connection between the canonical solitons and the
so-called nonautonomous ones. The finding of this transformation
has a significant contribution to understanding the essential
properties of the nonautonomous solitions and the dynamics of the
Bose-Einstein condensates by using the Feshbach resonance
technique.
\end{abstract}
\end{frontmatter}

\section{Introduction}

Dilute-gas Bose-Einstein condensates (BECs) have been generated by
many experiment groups since 1995 \cite{ande95,dalf99}. It is
well-known that the mean-field dynamics of the BECs at low
temperature can be well described by a three-dimensional
Gross-Pitaevskii (GP) equation \cite{gros,pita}, where the BECs
are confined to an expulsive harmonic external potential. In some
physically important cases the GP equation can be reduced
effectively to a one-dimensional GP equation
\cite{pel,kev,abdu2003,abdu2004,bra,hid,Liang,yuce}. This
simplified equation reads after re-scaling the physical variables,
\begin{equation}
i\hslash \frac{\partial }{\partial t}\psi (x,t)=\left(
-\frac{\hslash ^{2}}{ 2m}\frac{\partial ^{2}}{\partial
x^{2}}+g_{0}\Gamma (t) |\psi (x,t)|^{2}+V(x)\right) \psi (x,t),
\label{1}
\end{equation}
where $\psi (x,t)$ denotes the macroscopic wave function of the
condensate and $V(x)$ is the potential confining the condensate.
The $g_0$ represents the nonlinear interaction strength and a
time-dependent factor of $\Gamma (t)$ is controlled by the
so-called Feshbach resonance\cite{rob,ste}. It is very interesting
to study the dynamics of the BECs described by Eq. (\ref{1}) with
the Feshbach resonance management.

It should be pointed out that Eq. (\ref{1}) is a special case of a
more generalized nonlinear Schr\"odinger (NLS) equation with
varying in time and space dispersion and nonlinearity. In the
literature, this generalized NLS equation was named as the
nonautonomous NLS equation \cite{non-nlse}. The exact solutions of
the nonautonomous NLS equation are suggested to name the
nonautonomous solitons in order to conceptionally distinguish from
the canonical solitions introduced by Zabusky and Kruskal
\cite{can-soliton}. Actually, the studies of the integrability of
the generalized NLS equation have a long history. As early as in
1976 Chen and Liu \cite{chen76} went beyond the concept of the
canonical soliton and found the exact integrability and the Lax
pair of NLS equation with space-varying potentials. More recently,
the dynamics of the solitons of these generalized NLS equation has
been extensively studied \cite{non-nlse, dyna, non-soliton} and
some novel concepts, i.e., the nonautomonous NLS equation
\cite{non-nlse} and the nonautomonous soliton \cite{non-soliton}
have been proposed.

A classical way to study the integrability of the standard NLS
equation and the nonautomonous one is the construction of the Lax
pair of these NLS equations based on the inverse scattering
transform \cite{ist}. In this paper, we use the WTC test method
suggested by Weiss, Tabor and Carnevale \cite{wtc,as1} in 1983 to
study the integrability of the GP equation with Feshbach resonance
management. This method is based on the Painlev\'{e} test for PDE.
Through the WTC test, we get an integrability condition. At the
same time, we also find a transformation that can convert the GP
equation with Feshbach resonance management into the standard NLS
equation, which has been extensively studied and a number of exact
solutions including various solitons have been obtained in the
literature. From the well-known solutions of the standard NLS
equation, we can obtain novel solutions of the integrable GP
equation with Feshbach resonance management. This transformation
builds a systematical connection between the well-known canonical
soltions and the nonautomonous ones and provide a novel way to
study the dynamics of the BEC described by the GP equation with
Feshbach resonance management. Finally, we also find a new type of
solution of the standard NLS equation, which is also a new type of
solution of the GP equation.

The outline of this paper is as follows. In Sec. 2 we explicitly
present the WTC test of the GP equation (\ref{1}) and discuss the
condition the equation can pass the WTC test, i.e., the
integrability condition of Eq. (\ref{1}). In Sec. 3 we show a
transformation to convert the GP equation (\ref{1}) into the
well-known standard NLS equation, which demonstrates the complete
integrability of the GP equation under the condition obtained in
the previous section. In Sec. 4, as some applications, we list
some solutions of the standard NLS equation, including a new type
of solution, from which numerous novel solutions of the integrable
GP equation with Feshbach resonance management can be obtained by
the transformation.

\section{The Painlev\'{e} test for the GP equation with
Feshbach resonance management}

For simplicity, we rewrite Eq. (\ref{1}) as
\begin{equation}
i{\frac{\partial }{\partial t}}u\left( x,t\right) +{\frac{\partial
^{2}}{
\partial {x}^{2}}}u\left( x,t\right) +g\left( x,t\right) \left\vert u\left(
x,t\right) \right\vert ^{2}u\left( x,t\right) +V_{{0}}{x}^{2}u\left(
x,t\right) =0  \label{2}
\end{equation}
In order to use the Painlev\'{e} analysis, it is convenient to
introduce a complex function $v(x,t)=u(x,t)^{\ast }$. Thus,
Eq.(\ref{2}) becomes a pair of complex equations,
\begin{equation}
\left\{
\begin{array}{c}
i{\frac{\partial }{\partial t}}u\left( x,t\right) +{\frac{\partial
^{2}}{
\partial {x}^{2}}}u\left( x,t\right) +g\left( x,t\right) \left( u\left(
x,t\right) \right) ^{2}v\left( x,t\right) +V_{{0}}{x}^{2}u\left( x,t\right)
=0, \\
-i{\frac{\partial }{\partial t}}v\left( x,t\right)
+{\frac{\partial ^{2}}{
\partial {x}^{2}}}v\left( x,t\right) +h\left( x,t\right) \left( v\left(
x,t\right) \right) ^{2}u\left( x,t\right) +V_{{0}}{x}^{2}v\left(
x,t\right) =0,
\end{array}
\right.   \label{3}
\end{equation}
where $h(x,t)=g(x,t)^{\ast }$ .

The next step is to seek a solution of Eqs.(3) in the following
form
\begin{equation}
\left\{
\begin{array}{c}
u\left( x,t\right) =\phi (x,t)^{-p}\sum_{i=0}^{\infty }u_{{i}}\left(
t\right) \phi (x,t)^{i}, \\
v\left( x,t\right) =\phi (x,t)^{-q}\sum_{j=0}^{\infty
}v_{{j}}(t)\phi (x,t)^{j}.
\end{array}
\right.   \label{4}
\end{equation}
The coefficients $g(x,t)$ and $h(x,t)$ can be expanded on the
singularity manifold as follows
\begin{equation}
\left\{
\begin{array}{c}
g\left( x,t\right) =\sum_{k=0}^{\infty }g_{{k}}\left( t\right) \phi
(x,t)^{k}, \\
h\left( x,t\right) =\sum_{m=0}^{\infty }h_{{m}}\left( t\right)
\phi (x,t)^{m}.
\end{array}
\right.   \label{5}
\end{equation}
Here we use the Kruskal assumption $\phi(x,t)=x+\psi(t)$, where $
u_{i}(t),v_{j}(t),g_{k}(t),h_{m}(t)$ are analytic functions in the
neighborhood of a non-characteristic singularity manifold defined
by $x=-\psi (t)$.

By the standard procedure, the leading order is given by $p=q=1$,
and one can obtain the following relations
\begin{equation}
\left[
\begin{array}{cc}
Q_{{1}} & g_{{0}}\left( t\right) \left( u_{{0}}\left( t\right) \right) ^{2}
\\
\noalign{\medskip}h_{{0}}\left( t\right) \left( v_{{0}}\left(
t\right) \right) ^{2} & Q_{{2}}
\end{array}
\right] \left[
\begin{array}{c}
u_{{j}}\left( t\right)  \\
\noalign{\medskip}v_{{j}}\left( t\right)
\end{array}
\right] =\left[
\begin{array}{c}
F_{{j}} \\
\noalign{\medskip}G_{{j}}
\end{array}
\right] ,  \label{6}
\end{equation}
where $j=1,2,\cdots $, and
\begin{equation*}
\left\{
\begin{array}{c}
Q_{{1}}=\left( j-1\right) \left( j-2\right) +2\,g_{{0}}\left(
t\right) u_{{0}
}\left( t\right) v_{{0}}\left( t\right) , \\
Q_{{2}}=\left( j-1\right) \left( j-2\right) +2\,h_{{0}}\left(
t\right) u_{{0} }\left( t\right) v_{{0}}\left( t\right).
\end{array}
\right.
\end{equation*}
Here $g_{0}(t)$, $h_{0}(t)$, $u_{0}(t)$ and $v_{0}(t)$ obey
\begin{equation}
g_{{0}}\left( t\right) u_{{0}}\left( t\right) v_{{0}}\left(
t\right) =h_{{0} }\left( t\right) u_{{0}}\left( t\right)
v_{{0}}\left( t\right) =-2. \label{7}
\end{equation}
The recursion relations are determined by
\begin{eqnarray}
F_{{j}} &=&-{\frac{d}{dt}}u_{{j-2}}\left( t\right)
-iu_{{j-1}}\left( t\right) \left( {\frac{d}{dt}}\psi \left(
t\right) \right) j+2\,iu_{{j-1} }\left( t\right)
{\frac{d}{dt}}\psi \left( t\right) -g_{{j}}\left( t\right)
\left( u_{{0}}\left( t\right) \right) ^{2}v_{{0}}\left( t\right)   \notag \\
&&-g_{{0}}\left( t\right) v_{{0}}\left( t\right)
\sum_{m=1}^{j-1}u_{{j-m} }\left( t\right) u_{{m}}\left( t\right)
-g_{{0}}\left( t\right) \sum_{m=1}^{j-1}v_{{j-m}}\left( t\right)
\sum_{k=0}^{m}u_{{m-k}}\left(
t\right) u_{{k}}\left( t\right)   \notag \\
&&-\sum_{m=1}^{j-1}g_{{j-m}}\left( t\right)
\sum_{l=0}^{m}v_{{m-l}}\left( t\right)
\sum_{k=0}^{l}u_{{l-k}}\left( t\right) u_{{k}}\left( t\right)
-V_{{0
}}{x}^{2}u_{{j-2}}\left( t\right) ,  \label{8} \\
G_{j} &=&-{\frac{d}{dt}}v_{{j-2}}\left( t\right) +iv_{{j-1}}\left( t\right)
\left( {\frac{d}{dt}}\psi \left( t\right) \right) j-2\,iv_{{j-1}}\left(
t\right) {\frac{d}{dt}}\psi \left( t\right) -h_{{j}}\left( t\right) \left(
v_{{0}}\left( t\right) \right) ^{2}u_{{0}}\left( t\right)   \notag \\
&&-h_{{0}}\left( t\right) u_{{0}}\left( t\right)
\sum_{m=1}^{j-1}v_{{j-m} }\left( t\right) v_{{m}}\left( t\right)
-h_{{0}}\left( t\right) \sum_{m=1}^{j-1}u_{{j-m}}\left( t\right)
\sum_{k=0}^{m}v_{{m-k}}\left(
t\right) v_{{k}}\left( t\right)   \notag \\
&&-\sum_{m=1}^{j-1}h_{{j-m}}\left( t\right)
\sum_{l=0}^{m}u_{{m-l}}\left( t\right)
\sum_{k=0}^{l}v_{{l-k}}\left( t\right) v_{{k}}\left( t\right)
-V_{{0 }}{x}^{2}v_{{j-2}}\left( t\right) .  \label{9}
\end{eqnarray}
One notes that $g_j(t) = 0$ if $j < 0$. It is also true for
$h_j(t), u_j(t), v_j(t)$. The expressions $F_{j}$, $G_{j}$ for a
given $j $ depend only on the expansion coefficients of $u_{l}$,
$v_{l}$, with $l<j$ and $g_{l}$, $h_{l}$ with $l \leq j$.
Therefore the above equations represent the recursion relations of
the unknown $u_{j}$, $v_{j}$ ($j > l$) from the known $u_{l}$, $
v_{l}$ ( $l<j$).

The above recursion relations determine the unknown expansion
coefficients uniquely unless the determinant of the matrix in Eq.
(\ref{6}) is zero. Those values of j at which the determinant is
equal to zero are called as the resonance points. After some
calculations it is found that the resonance points only occur at
\begin{equation*}
j=-1,0,3,4.
\end{equation*}
The resonance point of $j=-1$ corresponds to the arbitrariness of
the singular manifold $\phi (x,t)$.

From the recursion relations, we find the compatibility conditions
for the remaining resonance points
\begin{eqnarray}
&&j=0:\,\,g_{{0}}\left( t\right) u_{{0}}\left( t\right) v_{{0}}\left(
t\right) =h_{{0}}\left( t\right) u_{{0}}\left( t\right) v_{{0}}\left(
t\right) =-2,  \label{10} \\
&&j=3:\,\,v_{{0}}\left( t\right) F_{{3}}=u_{{0}}\left( t\right) G_{{3}},
\label{11} \\
&&j=4:\,\,v_{{0}}\left( t\right) F_{{4}}+u_{{0}}\left( t\right) G_{{4}}=0.
\label{12}
\end{eqnarray}

Due to the arbitrariness of $u_{0}(t)$ and $v_{0}(t)$, Eq.
(\ref{10}) implies $g(x,t)\equiv h(x,t)$. According to the
definition of $h(x,t)$, one concludes that $g(x,t)$ must be a real
function.

According to Eq. (\ref{6}), one can obtain uniquely
$u_{1}(t),v_{1}(t),u_{2}(t),v_{2}(t)$. Inserting them into Eq.
(\ref{11}), after some manipulations one has
\begin{equation}
g_{{0}}(t){\frac{d}{dt}}g_{1}(t)-2\,g_{1}(t){\frac{d}{dt}}g_{0}(t)=0.
\label{13}
\end{equation}
Solving for $g_{1}(t)$, we obtain
\begin{equation}
g_{1}\left( t\right) =C_{{1}}\left( g_{{0}}\left( t\right) \right)
^{2}, \label{14}
\end{equation}
where $C_{1}$ is an arbitrary constant.

Inserting Eq. (\ref{14}) into Eq. (\ref{11}), one can obtain
$u_{3}(t)$ when $ v_{3}(t)$ is arbitrary. Replacing $u_{1}(t)$,
$v_{1}(t)$, $u_{2}(t)$, $ v_{2}(t)$, $u_{3}(t)$ into Eq.
(\ref{12}), after some simplifications and let the coefficients of
$\psi (t)$, ${\frac{d}{d{t}}}\psi (t)$, ${
\frac{d^{2}}{d{t}^{2}}}\psi (t)$ be zero (due to the arbitrariness
of $\psi (t)$), one has
\begin{equation}
C_{1}=0.  \label{15}
\end{equation}
Thus Eq. (\ref{14}) means that
\begin{equation}
g_{1}(t)=0.  \label{16}
\end{equation}

According to the definition of $g_{1}(t)$,
$g_{1}(t)={\frac{\partial }{\partial x}} g\left( x,t\right)
|_{x=-\psi (t)}$, when $\psi (t)$ is arbitrary, $g(x,t)$ must be
independent of $x$, so one has
\begin{equation}
\left\{
\begin{array}{l}
g(x,t)=g(t)=g_{0}(t), \\
g_{1}(t)=g_{2}(t)=\cdots =0.
\end{array}
\right.   \label{17}
\end{equation}
Finally, the conditions (\ref{15})-(\ref{17}), together with
compatibility condition (\ref{12}) give the following equation
\begin{equation}
4\,V_{{0}}\left( g_{{0}}\left( t\right) \right) ^{2}+\left(
{\frac{d^{2}}{d{t }^{2}}}g_{{0}}\left( t\right) \right)
g_{{0}}\left( t\right) -2\,\left( { \frac{d}{dt}}g_{{0}}\left(
t\right) \right) ^{2}=0.  \label{18}
\end{equation}
From this equation we have
\begin{equation*}
g_{{0}}\left( t\right) ={\frac{{e^{\pm 2\sqrt{V_{0}}t}}}{{A e^{\pm
4\sqrt{V_{{ 0}}}t}}-B}},
\end{equation*}
where $A,B$ are arbitrary constants. For a proper choice of $A$
and $B$, $g_0(t)$ can change its sign at certain time (singularity
point), which should be related to the Feshbach resonance.

Thus, one can conclude that only when
\begin{equation*}
g(x,t)=g(t)={\frac{{e^{\pm 2\,\sqrt{V_{{0}}}t}}}{{Ae^{\pm 4\sqrt{V_{0}}t}}-B}
},
\end{equation*}
Eq. (\ref{2}) can pass the WTC test. We will show below that this
condition is sufficient for the integrability of Eq. (\ref{2}).
For convenience, it is also useful to introduce $\lambda =\pm
2\sqrt{V_{{0}}}$, i.e., $V_{0}=\frac{ 1}{4}\lambda ^{2}$. Thus
$g(x,t)$ becomes
\begin{equation}
g(x,t)={\frac{2\,g_{0}{e^{\lambda t}}}{{A\,e^{2\lambda t}}-B}}.  \label{19}
\end{equation}

It is very interest to note that under the present NLS equation,
Eq. (\ref{19}) is completely satisfied the exact integrability
condition proposed by Serkin, Hasegawa, and Belyaeva [ see Eq. (2)
in Ref. \cite{non-soliton}]. However, from the present analysis,
Eq. (\ref{19}) is a necessary condition for the Eq. (\ref{2}) to
pass the WTC test. A sufficient integrability condition should be
shown by further exploring a transformation which converts the GP
equation (\ref{2}) into a standard integrable NLS equation.

\section{Integrability}

Under the condition Eq. (\ref{19}),  Eq. (\ref{2}) can be
rewritten as
\begin{equation}
i{\frac{\partial }{\partial t}}u\left( x,t\right) +{\frac{\partial
^{2}}{
\partial {x}^{2}}}u\left( x,t\right) +{\frac{2\,g_{0}{e^{\lambda t}}}{{
A\,e^{2\lambda t}}-B}}\left\vert u\left( x,t\right) \right\vert
^{2}u\left( x,t\right) +\frac{1}{4}\lambda ^{2}{x}^{2}u\left(
x,t\right) =0.  \label{20}
\end{equation}

It should be noted that when $A=0,$ $B=-1$ and $\lambda
>0$, Eq. (\ref{20}) is reduced to the case discussed in
Ref. \cite{Liang}.

Now we further prove that Eq. (\ref{20}) is completely integrable.
Our idea is to find a transformation to convert exactly Eq.
(\ref{20}) into the standard NLS equation
\begin{equation}
i\frac{\partial }{\partial T}Q(X,T)+\varepsilon \frac{\partial
^{2}}{
\partial X^{2}}Q(X,T)+\delta \left\vert Q(X,T)\right\vert
^{2}Q(X,T)=0.
\label{21}
\end{equation}
In the literature, one often takes $\varepsilon =\pm
\frac{1}{2},\delta =\pm 1$ or $\varepsilon =\pm 1,\delta =\pm 2$.

In order to reduce Eq. (\ref{20}) to the standard NLS equation
(\ref{21}), we look for a transformation of the form
\begin{equation}
\psi (x,t)=Q\left( X\left( x,t\right) ,T\left( t\right) \right)
e^{ia(x,t)+c(t)}. \label{22}
\end{equation}

One also notes that other similarity transformations
\cite{konotop2006, konotop2008} have been explored to reduce the
nonautomonous NLS equation to the standard NLS equation. In
comparison to Eq. (\ref{22}), the form of the similarity
transformation in Ref. \cite{konotop2006} is quite different from
ours. The transformation parameters in Ref. \cite{konotop2006} is
independent of the space variables and the space dependence of the
transformation is completely specialized by the explicit form of
the similarity transformation introduced. In Ref.
\cite{konotop2008}, another similarity transformation is
introduced. However, this transformation reduces the nonautomonous
NLS equation to a stationary NLS equation, as shown in Eq. (2) in
Ref. \cite{konotop2008}.

After some tedious and technical calculations, one can obtain the
explicit forms of the transformation \ref{22}. In the following we
present them in different cases.

\textbf{Case I:}$A = 0, B = -1$

In this case, Eq. (\ref{20}) becomes
\begin{equation}
i{\frac{\partial }{\partial t}}u\left( x,t\right) +
{\frac{\partial ^{2}}{
\partial {x}^{2}}}u\left( x,t\right) +2\,g_{0}{e^{\lambda t}}\left\vert
u\left( x,t\right) \right\vert ^{2}u\left( x,t\right)
+\frac{1}{4}\lambda ^{2}{x}^{2}u\left( x,t\right) =0,  \label{23}
\end{equation}
and the transformation reads
\begin{equation*}
\left\{
\begin{array}{l}
X\left( x,t\right) =\frac{2 g_{0} \varepsilon}{\delta}e^{\lambda
t}\left( x - \frac{2g_{0} \varepsilon e^{\lambda t}}{\delta \lambda}\right) , \\
T\left( t\right) =\frac{2\varepsilon g_{0}^{2}}{\delta ^{2}\lambda
} \left(e^{2\lambda t}-1\right), \\
a(x,t)=-\frac{\lambda}{4}\left(x^{2}-\frac{8\varepsilon g_{0}
e^{\lambda t}}{\delta \lambda} x + \frac{8\varepsilon
^{2}g_{0}^{2}e^{2\lambda t}}{\delta^2 \lambda^2}\right), \\
c(t)=\frac{\lambda }{2}\left( t-\frac{1}{\lambda} \ln \frac{\delta
}{2g_{0}\varepsilon  } \right).
\end{array}
\right.
\end{equation*}
Using this transformation, it is straightforward to check that Eq.
(\ref{23}) has the form of Eq. (\ref{21}). Thus, according to a
known solution $Q\left( X,T\right) $ of Eq. (\ref{21}), one can
obtain a solution of Eq. (\ref{23}) as follows
\begin{equation}
u(x,t)={\scriptstyle\sqrt{\frac{2\varepsilon g_{0}}{\delta
}}{e}^{\frac{1}{2}\lambda t}Q\left( \frac{2g_{0}\varepsilon
}{\delta }e^{\lambda t} \left(x - \frac{
2g_{0}\varepsilon}{\delta\lambda }e^{\lambda
t}\right),\frac{2\varepsilon g_{0}^{2}}{\delta ^{2}\lambda }
\left(e^{2\lambda t}-1\right)\right)}
{e^{-\frac{i\lambda}{4}\left(x^{2}-\frac{8\varepsilon g_{0}
e^{\lambda t}}{\delta \lambda} x + \frac{8\varepsilon
^{2}g_{0}^{2}e^{2\lambda t}}{\delta^2 \lambda^2}\right)} }.
\label{24}
\end{equation}

As noted in Ref. \cite{lara2005}, this solution has some
interesting features since it can take into account the fallout of
particles form the BEC for $\lambda < 0$ and the amplification of
a soliton if $\lambda > 0$. To show this, one can define a new
variable $Q'(x, t) = \sqrt{2g_0} e^{\frac12 \lambda t} u(x, t)$,
which obeys the equation
\begin{equation}
i\frac{\partial}{\partial t} Q'{x, t} + {\frac{\partial ^{2}}{
\partial {x}^{2}}}Q'(x,t) + |Q'(x,t)|^2Q'(x,t)
+\frac{1}{4}\lambda ^{2}{x}^{2}Q'(x,t) = \frac{i}{2} \lambda Q'(x,
t), \label{25}
\end{equation}
which includes an additional complex term at the right-hand side
of Eq. (\ref{25}).

One notes that only when $\varepsilon \delta g_{0}>0$, the above
transformation is well-defined. In particular, for $g_{0}>0$
$(g_{0}<0)$, $\varepsilon = 1,\delta = 2$ $(\delta = -2)$ or
$\varepsilon =\frac{1}{2},\delta =1$ $ (\delta =-1)$ Eq.
(\ref{23}) denotes the standard NLS equation, which has been
extensively studied in the literature.

\textbf{Remark:} From the solution (\ref{24}), it is obvious that
if $\lambda>0$, the solutions of Eq. (\ref{23}) can be compressed
into very high local matter densities by increasing the absolute
value of atomic scattering length. For the bright soliton
($g_{0}>0$, $\varepsilon \delta >0$) case, it has been discussed
in details in Ref. \cite{Liang}. For the dark soliton ($g_{0}<0$,
$\varepsilon \delta <0$) case, the situation is similar. In
addition, if $\lambda <0$, the solutions can be expanded into very
low local matter densities by decreasing the absolute value of
atomic scattering length, as clearly seen from Eq. (\ref{25}).

\textbf{Case II:} $A\neq 0$

In this case, the transformation can be obtained as follows
\begin{equation*}
\left\{
\begin{array}{l}
X\left( x,t\right) = \frac{2g_0 \varepsilon e^{\lambda
t}}{\delta\left( {e^{2\lambda t}}A-B \right)}\left(x - \frac{2g_0
\varepsilon}{\delta \lambda^2 A e^{\lambda t}}\right), \\
T\left( t\right) = - \frac{2{g_{{0}}}^{2}\varepsilon}{\lambda
{\delta } ^{2}A } \left( \frac{1}{ {e^{2\lambda \,t}}A-B } -
\frac{1}{ A-B } \right), \\
a(x,t) = {\frac{\lambda \left({A}{e^{2 \lambda t}} + B\right)
{x}^{2} + \frac{8 g_{{0}}\varepsilon }{\lambda \delta}e^{\lambda
t} x + \frac{8g^2_0
\varepsilon^2}{\lambda^{3} \delta^{2} A}} {4\left( {e^{2 \lambda t}}A-B\right) }}, \\
c(t)=\frac{1}{2}\lambda \left(t-\frac{1}{\lambda}\ln {\frac{\delta
\left( { e^{2\,\lambda \,t}}A-B\right) }{2g_0\varepsilon }
}\right).
\end{array}
\right.
\end{equation*}
Likewise, this transformation can convert Eq. (\ref{20}) into the
form of Eq. (\ref{21}). Note that only when
$$\frac{\delta\,({e^{2\,\lambda \,t}}A-B) }{ g_{{0}}\varepsilon }>0,$$ the above
transformation is well-defined. The sign of $\varepsilon \delta $ is
determined by the signs of $\lambda $ and $g_{0}$, and the values of
$A$ and $B$.

\section{Explicit solutions of the GP equation}

As pointed out above, the solutions of the GP equation (\ref{20})
can be obtained from the solutions of the standard NLS equation
(\ref{21}). This is a very efficient method to find new solutions
of the GP equation (\ref{20}) since a number of solutions of the
standard NLS equation (\ref{21}) have been obtained by some
classical methods such as the inverse scattering transform, the
Lax pairs, the bilinear technique, and the Backlund
transformation, and so on.

In the following we first list some well-known solutions of the
standard NLS equation (\ref{21}), which can be transformed into
the solutions of the GP equation (\ref{20}). As an example, we
explicitly present the fundamental bright one-soliton solution of
the GP equation.

\textbf{Fundamental bright one-soliton solution}
\begin{equation*}
Q(X,T)={\it sech}(X)e^{\frac{i}{2}T}
\end{equation*}
when $g_{0}>0,$ $\varepsilon =\frac{1}{2}, \delta =1$. According
to Eq. (\ref{24}), the fundamental bright one-soliton solution of
the GP equation (\ref{22}) can be written as
\begin{eqnarray*}
u(x,t) && = \sqrt{g_0}e^{\frac12\lambda t}Q\left[ g_{0}e^{\lambda
t}\left(x - \frac{g_{0}}{\lambda }e^{\lambda
t}\right),\frac{g_{0}^{2}}{\lambda }\left(e^{2\lambda t} -
1\right)\right]} { e^{-\frac{i\lambda}{4}\left( x^2-\frac{4 g_0
}{\lambda} e^{\lambda t} x + \frac{2g^2_0}{\lambda^2}e^{2\lambda
t}\right)} \\
&& = \sqrt{g_0} e^{\frac12\lambda t} sech \left[g_{0}e^{\lambda
t}\left(x - \frac{g_{0}}{\lambda }e^{\lambda t}\right)\right]
e^{-\frac{i\lambda}{4}\left( x^2-\frac{4 g_0 }{\lambda} e^{\lambda
t} x +\frac{2g^2_0}{\lambda^2}e^{2\lambda t}\right)}.
\end{eqnarray*}

For simplicity, below we only list the soliton solutions of the
standard NLS equation, the corresponding solutions of the GP
equation (\ref{20}) can be obtained by the same procedure shown
above.

\textbf{Fundamental dark one-soliton solution}
\begin{equation*}
Q(X,T)=q_{{0}}\tanh (q_{{0}}X)e^{iq_{{0}}^{2}T}
\end{equation*}
when $q_{{0}}$ is an arbitrary constant and $ g_{0}<0,$
$\varepsilon =-\frac{1}{2},\delta =1$.

\textbf{Envelope solutions of the bright soliton}
\begin{equation*}
Q(X,T)=k_{1}\mathit{sech}\left( k_{{1}}X-\,k_{{1}}k_{{2}}T\right) {\
e^{-i\left( -k_{{2}}X+\frac{1}{2}\left( {k_{{2}}}^{2}-{k_{{1}}}^{2}\right)
T\right) }.}
\end{equation*}
where $k_{{1}},$ $k_{{2}}$ are arbitrary constants, and
$\varepsilon =\frac{ 1 }{2},\delta =1$.

\textbf{Envelope solutions of the dark soliton}
\begin{equation*}
Q(X,T)=k_{1}\mathit{\tanh }\left( k_{{1}}X+\,k_{{1}}k_{{2}}T\right) {\
e^{-i\left( -k_{{2}}X-\frac{1}{2}\left( {k_{{2}}}^{2}+2{k_{{1}}}^{2}\right)
T\right) }.}
\end{equation*}
where $k_{{1}},$ $k_{{2}}$ are arbitrary constants, and $\varepsilon =-\frac{
1}{2},\delta =1$.

For bright and dark N-soliton solutions of Eq. (\ref{21}), one can
refer to Refs.\cite{as2,zs1,zs2,sy,nm,blow,mlb}.

Besides the well-known soliton solutions listed above, in the
following we discuss a new type exact solution of the standard NLS
equation (\ref{21}) by using the direct truncation method proposed
in Ref. \cite{zlw}. These new solutions, of course, also give new
solutions of the GP equation (\ref{20})

Assume that
\begin{equation*}
Q(X,T)=\eta (k_{1}X+\omega _{1}T)e^{i(k_{2}X+\omega _{2})}
\end{equation*}
be a solution of Eq.(\ref{21}), where $\eta (.)$ is a real
function. Some straightforward calculations give
\begin{equation*}
\omega_1=-2\,\varepsilon \,k_{{1}}k_{{2}},
\end{equation*}
and
\begin{equation}
-\varepsilon \eta \left( \xi \right) {k_{{2}}}^{2}-\eta \left( \xi
\right) \omega _{{2}}+\varepsilon \left( {\frac{d^{2}}{d{\xi
}^{2}}}\eta \left( \xi \right) \right) {k_{{1}}}^{2}+\delta
\,\left( \eta \left( \xi \right) \right) ^{3}=0,  \label{26}
\end{equation}
where $\xi =k_{1}X+\omega _{1}T$. By the similar procedure
proposed in Ref. \cite {zlw} we set
\begin{equation}
\eta \left( \xi \right) =\frac{a+bf(\xi )+cg(\xi )}{p+qf(\xi
)+rg(\xi )}, \label{27}
\end{equation}
where $a, b, c, p, q, r$ are constants to be determined together
with $k_{{1}},k_{ {2}}$ and $\omega _{2}$. The functions $f(\xi )$
and $g(\xi )$ are assumed to obey
\begin{equation}
\left\{
\begin{array}{l}
f(\xi )^{2}+g(\xi )^{2}=1, \\
\frac{d}{d\xi }f(\xi )=g(\xi )\sqrt{1-k^{2}f(\xi )^{2}}, \\
\frac{d}{d\xi }g(\xi )=-f(\xi )\sqrt{1-k^{2}f(\xi )^{2}},
\end{array}
\right.   \label{28}
\end{equation}
Inserting (\ref{27}) and (\ref{28}) into Eq. (\ref{26}), we get a
polynomial of $f(\xi )$ and $g(\xi )$. Collecting the coefficients
and let them be zero, one obtains a set of algebraic equations of
$a$, $b$, $c$, $p$, $q$, $r$, $k_{{1}}$, $k_{{2}}$ and $\omega
_{{2}}$. Solving for the set of equations one can obtain many
exact solutions. Here we only present three representative
solutions.

{\bf The solution I:}
\begin{equation*}
Q(X,T)=\sqrt{\frac{2\varepsilon }{\delta }}kk_{1}cn\left( k_{{1}
}X-2\,\varepsilon \,k_{{1}}k_{{2}}T,k\right) {e^{-i\left( -k_{{2}
}X+\varepsilon \left( {k_{{1}}}^{2}+{k_{{2}}}^{2}-2\,{k_{{1}}}^{2}{k}
^{2}\right) T\right) },}
\end{equation*}
where $k_{{1}},k_{{2}}$ are arbitrary constants and
$\varepsilon\delta>0$. When $k\rightarrow 1,$ we obtain the envelope
solutions of bright soliton
\begin{equation*}
Q(X,T)=\sqrt{\frac{2\varepsilon }{\delta }}kk_{1}\mathit{sech}\left( k_{{1}
}X-2\,\varepsilon \,k_{{1}}k_{{2}}T\right) {e^{-i\left( -k_{{2}
}X+\varepsilon \left( {k_{{2}}}^{2}-{k_{{1}}}^{2}\right) T\right) }.}
\end{equation*}
When $k\rightarrow 0,$ we obtain the period solutions
\begin{equation*}
Q(X,T)=\sqrt{\frac{2\varepsilon }{\delta }}kk_{1}\mathit{\cos }\left( k_{{1}
}X-2\,\varepsilon \,k_{{1}}k_{{2}}T\right) {e^{-i\left( -k_{{2}
}X+\varepsilon \left( {k_{{1}}}^{2}+{k_{{2}}}^{2}\right) T\right) }.}
\end{equation*}

{\bf The solution II:}
\begin{equation*}
Q(X,T)=\sqrt{-\frac{2\varepsilon }{\delta }}kk_{1}sn\left( k_{{1}
}X-2\,\varepsilon \,k_{{1}}k_{{2}}T,k\right) {e^{-i\left( -k_{{2}
}X+\varepsilon \left( {k_{{1}}}^{2}+{k_{{2}}}^{2}+\,{k_{{1}}}^{2}{k}
^{2}\right) T\right) },}
\end{equation*}
where $k_{{1}},k_{{2}}$ are arbitrary constants and
$\varepsilon\delta<0$. When $k\rightarrow 1,$ we obtain the envelope
solutions of dark soliton
\begin{equation*}
Q(X,T)=\sqrt{-\frac{2\varepsilon }{\delta }}kk_{1}\mathit{\tanh
}\left( k_{{ 1 }}X-2\,\varepsilon \,k_{{1}}k_{{2}}T\right)
{e^{-i\left( -k_{{2} }X+\varepsilon \left(
{k_{{2}}}^{2}+2{k_{{1}}}^{2}\right) T\right) }.}
\end{equation*}
When $k\rightarrow 0,$ we obtain the period solutions
\begin{equation*}
Q(X,T)=\sqrt{-\frac{2\varepsilon }{\delta }}kk_{1}\mathit{\sin }\left( k_{{1}
}X-2\,\varepsilon \,k_{{1}}k_{{2}}T\right) {e^{-i\left( -k_{{2}
}X+\varepsilon \left( {k_{{1}}}^{2}+{k_{{2}}}^{2}\right) T\right) }.}
\end{equation*}

{\bf The solution III:}
\begin{equation*}
Q(X,T)=\sqrt{-\frac{\varepsilon }{2\delta }}k_{1}\frac{sn\left(
k_{{1} }X-2\,\varepsilon \,k_{{1}}k_{{2}}T,k\right) }{1+cn\left(
k_{{1} }X-2\,\varepsilon \,k_{{1}}k_{{2}}T,k\right)
}{e^{-\frac{i}{2}\left( -2k_{{2} }X+\varepsilon \left(
2{k_{{2}}}^{2}-{k_{{1}}}^{2}+\,2{k_{{1}}}^{2}{k} ^{2}\right)
T\right) }.}
\end{equation*}
where $k_{{1}},k_{{2}}$ are arbitrary constants. When
$k\rightarrow 1$, we obtain the envelope solutions
\begin{equation*}
Q(X,T)=\sqrt{-\frac{\varepsilon }{2\delta }}k_{1}\frac{\tanh \left( k_{{1}
}X-2\,\varepsilon \,k_{{1}}k_{{2}}T\right) }{1+\mathit{sech}\left( k_{{1}
}X-2\,\varepsilon \,k_{{1}}k_{{2}}T\right) }{e^{-\frac{i}{2}\left( -2k_{{2}
}X+\varepsilon \left( 2{k_{{2}}}^{2}+\,{k_{{1}}}^{2}\right) T\right) }.}
\end{equation*}

To our knowledge, the solution III is of a new type.

\textbf{Remark 1 }Set $k_{{2}}=0$ in the solutions I, II, and III,
we get the stationary solutions.

\textbf{Remark 2 }Obviously, when $\varepsilon \delta >0$, the
standard NLS equation has bright soliton solutions, and when $
\varepsilon \delta <0$, it has dark soliton solutions.

\textbf{Remark 3 } As mentioned above, only when $\varepsilon
\delta g_{0}>0$, the transformation is well-defined. According to
Remark 2 we know that, when $g_{0}>0$, the GP equation (\ref{23})
has bright soliton solutions and when $g_{0}<0$, the GP equation
(\ref{23}) has dark soliton solutions. For the GP equation
(\ref{20}), one can use the Feshbach resonance to control the
bright and the dark soliton in the BECs. This is because that if
$\frac{2\,g_{0}{e^{\lambda t} }}{{A\,e^{2\lambda t}}-B}>0$, Eq.
(\ref{20}) is corresponding to Eq. ( \ref{21}) with $\varepsilon
\delta>0$, and if $\frac{ 2\,g_{0}{e^{\lambda t}}}{{A\,e^{2\lambda
t}}-B}<0$, Eq. (\ref{20}) is corresponding to the case of
$\varepsilon \delta <0$. The exact solutions of the GP equation
with Feshbach resonance management presented in this paper have a
significant contribution to the experimental study of the dynamics
of the BECs.

In conclusion, we discuss the integrability of the GP equation with
Feshbach resonance management. By WTC test we find a condition under
which the GP equation is completely integrable. Meanwhile, we also
find a transformation, which can convert the GP equation with
Feshbach resonance management into the standard NLS equation. By
this transformation, the well-known exact solutions of the standard
NLS equation can be converted into the exact solutions of the GP
equation. To our knowledge, this is the first time to give a
correspondence from the exact solutions of the standard NLS equation
to the exact solutions of the GP equation, and thus we can give
abundant exact solutions for the GP equation. These solutions would
be important for understanding the dynamical behavior of the BECs by
using the Feshbach resonance technique.

\textbf{Acknowledgement} This work was supported by the National
Natural Science Foundation under the grant No.10205007 of China.

\end{document}